\documentclass[useAMS,usenatbib, letter]{mnras}
\usepackage[T1]{fontenc} 
\usepackage{aecompl}
\usepackage{comment}
\usepackage{amsmath}
\usepackage{amssymb}
\usepackage{graphicx, subfig}

\def\refjnl#1{{\rm#1}}

\def\counits{\rm K\,km\, s^{-1}\,pc^2}
\def\aj{\refjnl{AJ}}                   
\def\apj{\refjnl{ApJ}}                 
\def\apjl{\refjnl{ApJ}}                
\def\apjs{\refjnl{ApJS}}               
\def\ao{\refjnl{Appl.~Opt.}}           
\def\aap{\refjnl{A\&A}}                

\def\mnras{\refjnl{MNRAS}}             
\def\be{\begin{equation}} 
\def\ee{\end{equation}}

\def\gsim{\lower.5ex\hbox{\gtsima}} 
\def\lsim{\lower.5ex\hbox{\ltsima}} 
\def\gtsima{$\; \buildrel > \over \sim \;$} 
\def\ltsima{$\; \buildrel < \over \sim \;$} 
\def\prosima{$\; \buildrel \propto \over \sim \;$} \def\gsim{\lower.5ex\hbox{\gtsima}} 
\def\lsim{\lower.5ex\hbox{\ltsima}} 
\def\simgt{\lower.5ex\hbox{\gtsima}} 
\def\simlt{\lower.5ex\hbox{\ltsima}} 
\def\simpr{\lower.5ex\hbox{\prosima}} \def\la{\lsim} \def\ga{\gsim} 
  
\def\gtsima{$\; \buildrel > \over \sim \;$} 
\def\ltsima{$\; \buildrel < \over \sim \;$} 
\def\gsim{\lower.5ex\hbox{\gtsima}} 
\def\lsim{\lower.5ex\hbox{\ltsima}} 
\def\simgt{\lower.5ex\hbox{\gtsima}} 
\def\simlt{\lower.5ex\hbox{\ltsima}} 
\def\simpr{\lower.5ex\hbox{\prosima}} 
\def\la{\lsim} 
\def\ga{\gsim}

\def\E3{{\cal E}_{\rm g}^{III}}

\title[]{CO luminosity function from Herschel-selected galaxies and the contribution of AGN}
\author[Vallini et al.]{L. Vallini$^{1,2}$\thanks{E-mail: livia.vallini@unibo.it}, C. Gruppioni$^2$, F. Pozzi$^{1, 2}$, C. Vignali$^{1,2}$, G. Zamorani$^{2}$\\
$^1$ Dipartimento di Fisica e Astronomia, Universit\'a di Bologna, viale Berti Pichat 6/2, 40127 Bologna, Italy\\
$^2$ INAF - Osservatorio Astronomico di Bologna, via Ranzani 2, 40127, Bologna, Italy\\	
}

\begin{document}

\pagerange{\pageref{firstpage}--\pageref{lastpage}} \pubyear{2012}

\maketitle

\label{firstpage}

\begin{abstract}
We derive the CO luminosity function (LF) for different rotational transitions (i.e. (1--0), (3--2), (5--4)) starting from the \emph{Herschel} LF by Gruppioni et al. and using appropriate $L_{\rm CO} - L_{\rm IR}$ conversions for different galaxy classes. Our predicted LFs fit the data so far available at $z\approx0$ and $2$.
We compare our results with those obtained by semi-analytical models (SAMs): while we find a good agreement over the whole range of luminosities at $z\approx0$, at  $z\approx1$ and $z\approx2$ the tension between our LFs and SAMs in the faint and bright ends increases. We finally discuss the contribution of luminous AGN ($L_{X}>10^{44}\,\rm{erg\,s^{-1}}$) to the bright end of the CO LF concluding that they are too rare to reproduce the actual CO luminosity function at $z\approx2$.
\end{abstract}



\section{Introduction}\label{intro}

The study of the star formation (SF) history, and its connection with the gas mass accretion/consumption in ga\-laxies, is one of the still open issues in modern cosmology. A proper description of the evolution of SF across cosmic time needs both \emph{(a)} a thorough understanding of the relation between the total/molecular gas mass and the star formation, and \emph{(b)} sufficiently large samples of galaxies at different redshifts ($z$). 
As dust and gas are intimately associated, the dust infrared continuum emission can be a good proxy to infer the interstellar medium (ISM) mass \citep[][]{scoville2014, groves2015}, tracing it on large samples across cosmic time \citep[e.g.][]{berta2013}. The state-of-the-art Atacama Large (Sub)Millimeter Array (ALMA) will make it possible in the next future to directly follow the molecular gas abundance as a function of redshift with blind searches of carbon monoxide (CO) rotational transitions \citep[e.g][and references therein]{carilli2013}.
So far, only a handful of observational works have attempted to constrain this quantity. \citet{keres2003} measured for the first time the CO(1--0) luminosity function at $z=0$ u\-sing far-infrared (FIR) and optical B-band selected samples \citep[see also][for more recent CO(1--0) data at $z\approx0$]{boselli2014}. At $z\approx2$ we have some observational constraints by \citet{aravena2012} and \citet{daddi2010}. More recently, \citet{walter2014} measured the CO LF in three redshift bins ($z \approx 0.3, 1.52, 2.75$) based on a blind molecular line scan using the IRAM Plateau de Bure Interferometer. In the near future, the advent of similar searches with ALMA will enable similar studies to much deeper levels and over larger areas. 
On the theoretical side, the method generally adopted to predict CO ($J-(J-1)$)
LFs is to couple cosmological simulations with semi-analytical prescriptions that relate the CO emission to the physical properties of the simulated galaxies such as the intensity of the radiation field, the metallicity, the presence of an Active Galactic Nucleus (AGN) \citep[e.g.][]{obreschkow2009, lagos2012, fu2012, popping2014b}.
The aim of this letter is to derive the CO(1--0), CO(3--2), and CO(5--4)
luminosity functions at different redshifts by a\-dopting a simple empirical approach that allows to convert the state-of-the-art observed infrared LF presented in
\citet{gruppioni2013}. As a matter of fact, the CO luminosity is found
to correlate with the total infrared luminosity ($L_{\rm IR}$; $8-1000\,\rm{\mu m}$), providing
an integrated proxy of the Kennicutt-Schmidt \citep{kennicutt1998} relation that links star formation rate (SFR) and the molecular gas surface density. 
The correlation between these quantities relies on the fact that $L'_{\rm CO}$\footnote{In what follows the $L'_{\rm CO}$ notation will be used when the CO luminosity is expressed in $\counits$.} is a molecular hydrogen tracer, while $L_{\rm IR}$ is a proxy of the star formation rate.
This is true in homogeneous samples of galaxies with comparable interstellar medium properties, as the correlation between $L_{\rm CO}$ and $L_{\rm IR}$ implicitly depends on the dust-to-gas ratios and metallicity within the galaxies \citep[e.g.][]{leroy2013}, on the presence of additional heating due to AGN activity which affects the temperature of dust grains, and on the effects of gas streaming motions on the star-forming properties \citep[e.g.][]{meidt2013}.

\section{Method}
\begin{figure*}
\includegraphics[scale=0.40]{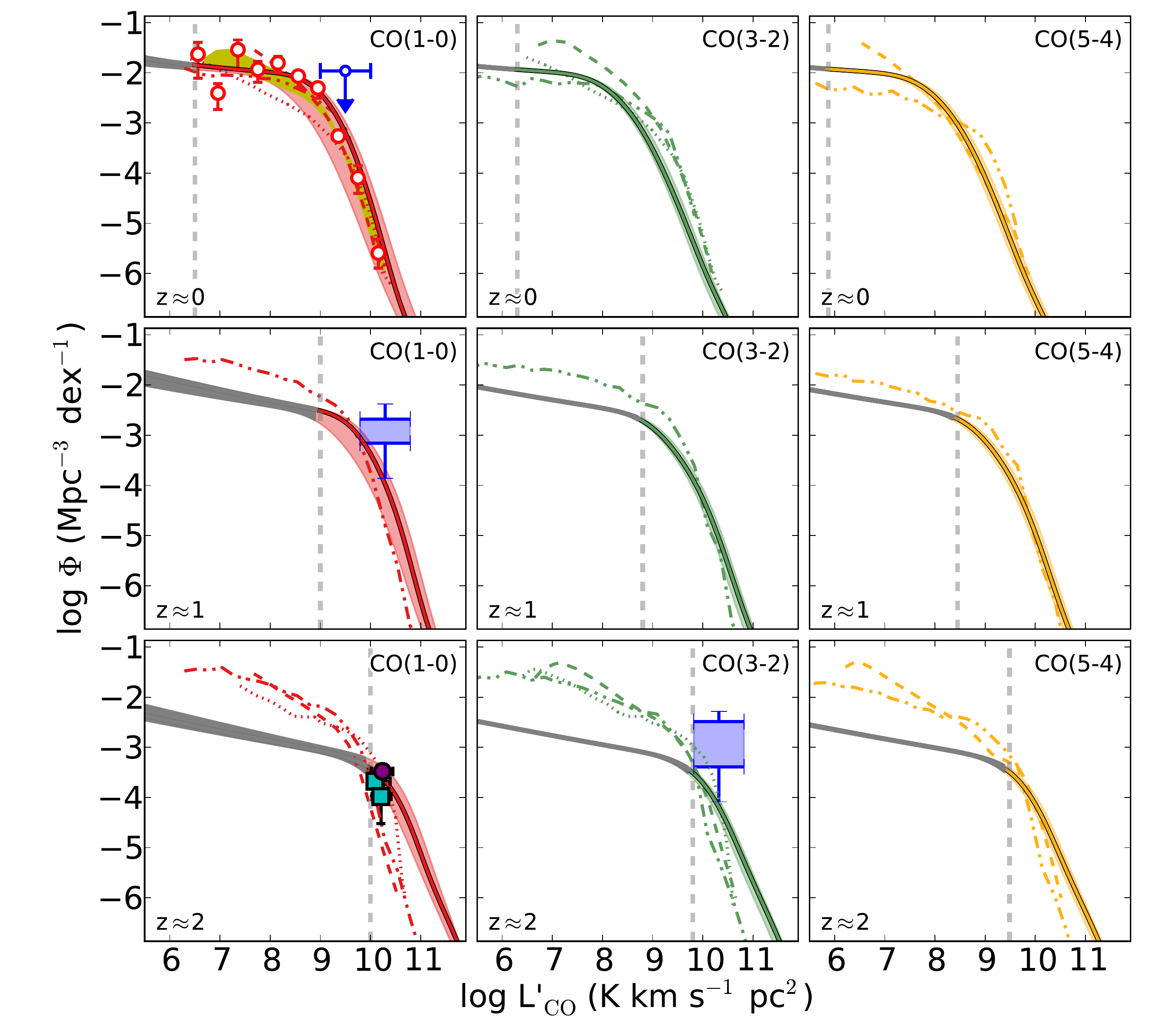}
\centering
\caption{From left to right: CO(1--0), CO(3--2), and CO(5--4) luminosity functions, at $z\approx0$ (upper row), $z\approx1$ (middle row), and $z\approx2$ (bottom row). Colored solid lines with shaded regions represent our prediction with the corresponding uncertainty. The faint-end extrapolation of the CO luminosity functions is plotted in gray and the luminosity limit of the underlying IR LFs is highlighted with a vertical dashed line.
Open circles at $z\approx0$ are the data from \citet{keres2003}, while cyan squares at $z\approx2$ are the space densities derived from CO detections presented in \citet{aravena2012} and corrected for the overdensity of the field in which the observation was performed. The purple circle in the same panel represents the data by \citet{daddi2010} from the BzK star-forming galaxies. The recent CO(1--0) and CO(3--2) data from \citet{walter2014} with their error bars are plotted in blue. Dashed, dotted, and dash-dotted lines are instead the predictions from semi-analytical models by \citet{obreschkow2009, lagos2012} and Popping et al. 2015 (in preparation) respectively. The yellow shaded region in the first panel highlights the parameter space covered by the three different models for the CO(1--0) LF at $z\approx0$ discussed by \citet{fu2012}.}\label{co_fig1}

\end{figure*}

We derive the CO LF starting from the IR LF \citep{gruppioni2013} and converting it into CO LF through empirical $L'_{\rm CO} - L_{IR}$ relations from the literature.
To this purpose, we consider the \citet{gruppioni2013} total IR luminosity function based on deep and extended far infrared ($70-500\,\rm{\mu m}$) data from the cosmological guaranteed time \emph{Herschel} surveys, PACS Evolutionary Probe, \citep[PEP; ][]{lutz2011} and Herschel Multi-tiered Extragalactic Survey, \citep[HerMES; ][]{oliver2012}, in the GOODS (GOODS-S and GOODS-N), Extended Chandra Deep Field South (ECDFS), and Cosmic Evolution Survey (COSMOS) areas. In \citet{gruppioni2013} the authors have completely characterized the multi-wavelength SEDs of the PEP sources by performing a detailed SED-fitting analysis and comparison with known template libraries of IR populations. The sources have been classified on the basis on their broad-band SEDs in five main classes: \texttt{spiral}, reproduced by templates of normal spiral galaxies, \texttt{starburst} reproduced by templates of starburst galaxies, \texttt{AGN1, AGN2}, reproduced by AGN-dominated SEDs (unobscured and obscured in the optical/UV), and \texttt{SF-AGN} reproduced by templates of Seyfert2/1.8/LINERS/Ultra Luminous Infrared Galaxies (ULIRGs) + AGN  (i.e., the AGN emission might be present, although not dominant). The latter class is further di\-vi\-ded into two sub-classes: \texttt{AGN-SB} and \texttt{AGN-GAL} on the basis of the far-IR/near-IR colours and on the evolutionary path. More precisely: the \texttt{AGN-SB} objects show an enhanced far-IR flux typical of starburst galaxies and dominate at high redshifts as the AGN-dominated sources, while \texttt{AGN-GAL} are characterized by a SED typical of normal spiral galaxies but with a low-luminosity AGN showing up in the mid-infrared.
For the shape of the LF, \citet{gruppioni2013} assumed a modified Schechter function \citep{saunders1990}, which depends on four parameters ($\alpha$, $\sigma$, $L^\star_{\rm IR}$ and $\Phi^\star$). The parameters $\alpha$ and $\sigma$ have
been estimated at the redshift where the corresponding LF of each population is best
sampled. This redshift is $z\approx0$ except for the \texttt{AGN1} and \texttt{AGN2} classes which are more numerous and whose LF is better defined at $z\approx2$. Subsequently, $\alpha$ and $\sigma$ have been frozen leaving only $L^\star$ and $\Phi^\star$ free
to vary. If we assume that the CO LF of \textit{each homogeneous class of objects} has the same $z$-evolution of the corresponding IR LF, we can calculate the evolution of the total CO LF by adopting physically motivated conversions between IR and CO luminosities according to the properties of the galaxies 
composing each class.
More precisely, $L_{\rm CO}$ and $L_{\rm IR}$ are generally found to be linked by a correlation of the form ${\rm log} (L'_{\rm CO}/\counits) = \alpha + \beta \, log (L_{\rm IR}/L_{\odot})$.\\
In the literature there are many studies regarding the CO-IR relation in different samples of galaxies ranging from starburst galaxies (SB)\citep[e.g.][]{greve2014} to normal (main sequence; MS) ones \citep[e.g][]{sargent2014, daddi2015}. 
According to the definition of \citet{rodighiero2011}, the SB galaxies are the objects that, at each redshift, are more than $\approx \,0.6\, \rm dex$ above the main sequence defined in the SFR-stellar mass plane. In the sample of \citet{gruppioni2013} only \texttt{AGN1, AGN2} and \texttt{AGN-SB} meet this criterium at all redshifts \citep[cfr. Fig. 15 of][]{gruppioni2013}; hence for these classes we will adopt the CO-IR relation found for starbursts by \citet{greve2014}.
In particular, \citet{greve2014} provide for the first time IR/FIR-CO luminosity relations that extend up to J$_{up}$=13 based on \emph{Herschel} SPIRE-FTS and ground-based telescopes data for local (U)LIRGs and high-$z$ sub-millimeter galaxies.  According to the notation introduced above, they found $\alpha \equiv \alpha_{SB}=[-2.0, -2.2, -2.9]$ and $\beta \equiv \beta_{SB}=[1.00, 1.00, 1.03]$ for the CO(1--0), CO(3--2), and CO(5--4) transitions respectively.

On the other hand, in \citet{gruppioni2013} galaxies with SEDs dominated by star formation (\texttt{spiral, starburst} and \texttt{AGN-GAL}), especially at $z>1.2$, are below the trashold defined by \citet{rodighiero2011} for starburst galaxies. Thus for these classes we will adopt the CO-IR conversion found by \citet{sargent2014} for MS galaxies.
More precisely, \citet{sargent2014}, considering a sample of $z\leq3$ MS galaxies with CO detections, found that the CO(1--0) luminosity correlates with the IR with parameters $\alpha \equiv \alpha_{MS} = 0.54 \pm 0.02$ and $\beta \equiv \beta_{MS}=0.81 \pm 0.03$. The $1\sigma$ scatter around the best fit relation is $\sigma_{MS}=0.21\,\rm{dex}$. 
For higher-J transitions we adopt the recent findings of \citet{daddi2015} in which it has been shown that the relation between $L_{\rm IR}$ and the CO(5--4) luminosity is well described by a linear relation with $\alpha^{54}_{MS}=-2.52$ and
$\beta^{54}_{MS}=1$. The dispersion in the residuals is $0.24\,\rm{dex}$. To convert the $L'_{\rm CO}$(5--4) into the
$L'_{\rm CO}$(3--2) we combine the average CO(3--2)/CO(1--0) flux ratio ($R_{31}=0.42 \pm 0.07$) and the CO(5--4)/CO(1--0) flux ratio ($R_{51}=0.23\pm 0.04$) measured by the
same authors within the same sample of galaxies. Although assuming conversion factors between different CO ($J-(J-1)$) transitions represents a strong assumption, our choice is motivated by the lack of explicit studies regarding the CO(3--2)-IR relation in MS galaxies. Ultimately, to obtain the total CO LF at various redshifts, we combine the luminosity function of each class after having  conveniently re-binned each of them within equal CO luminosity bins.

\section{Results}

In Figure \ref{co_fig1} we show with solid lines the CO(1--0), CO(3--2), and CO(5--4)
luminosity functions at $z\approx 0,\,1,\,2$ as obtained
through the method described in the previous Section. The LFs are color-coded
as a function of the transition, with shaded regions representing the
uncertainties on the prediction due to the scatter in the CO-IR relation. More precisely, the upper limit is obtained by adopting the conversion factors from \citet{sargent2014} and \citet{daddi2015} for all the populations and considering the maximum in their $L'_{\rm CO}-L_{\rm FIR}$ relation; the lower bound is obtained by considering the minimum in the $L'_{\rm CO}-L_{\rm FIR}$ relation from \citet{greve2014}. We find that our fiducial model is consistent with the observed points at $z=0$ \citep{keres2003} and the upper limit by \citet{walter2014}.
The same holds true for the CO(1--0) LF at $z\approx1$, even though it must be noticed that data point by \citet{walter2014} refers to a slightly higher redshift ($z\approx1.5$). The lower (upper) limits of the \citet{walter2014} data are obtained considering only secure detections (all candidates).
At $z\approx2$ the CO(1--0) observations by \citet{aravena2012} and \citet{daddi2015}, and the CO(3--2) data point at $z\approx2.7$ by \citet{walter2014} are well consistent with our LFs. However, it must be noticed that, on average, the estimate by \citet{walter2014} of the number densities is systematically higher with respect to our predictions.
We also compare our results with predictions of semi-analytical (SAM)
models by \citet{obreschkow2009, lagos2012, fu2012}, and with the CO LFs obtained by Popping et al. (in preparation) based on the model presented in \citet{popping2014b} where the authors coupled the \citet{popping2014a} semi-analytic model with a radiative transfer code.
While our results and SAMs are fairly in agreement over the whole range of luminosities at $z\approx0$, at  $z\approx1$ and $z\approx2$ they increasingly disagree at the faint and bright ends.
The difference in the faint end can
be explained by considering two concurrent reasons.
The first one is the well known excess of low/intermediate mass
galaxies predicted by most SAMs with respect to the observed
mass functions. The second one is related to the fact that the slope of the faint-end of the observed IR LFs is derived at $z\approx0$ where the LF is better covered by the data and kept fixed in all the redshift bins. This means that at $z\approx2$, where the faint
end is not covered by the infrared data (light gray regions in Fig.\ref{co_fig1}), the slope is not constrained by the observations. We note that, above this limit, our CO(1--0) luminosity function at $z\approx2$ reproduces the observed data from \citet{aravena2012}.
The overprediction of the bright-end at $z\approx2$ with respect to the SAMs shows up especially for the CO(1-0) transition that is the one more closely related to the star formation and scarcely affected by the AGN activity. Not surprisingly, the same trend has been recently found by \citet{gruppioni2015} when comparing the SFR function derived from IR luminosity (due to the SF only) with the SFR function predicted by four different SAMs. 
These discrepancies might be connected either to wrong photometric redshifts and source confusion that might enhance the bright-end of the \textit{Herschel} IR LF, or to the difficulty of SAMs in modeling the AGN feedback that affects the inflow/outflow of gas in the largest and most massive galaxies. We note that for high-J CO transitions the tension between our LFs and SAMs at $z\approx2$ is less obvious. However, high-J CO lines luminosities are strongly dependent on the CO Spectral Line Energy Distribution that varies from galaxy to galaxy. We have tested that if, instead of adopting the $L'_{CO}($J-(J-1)$)-IR$ relation provided by \citet{greve2014} to convert the IR LF of starburst galaxies, we consider the maximum and minimum $L'_{CO}(3-2)/L'_{CO}(1-0)$ ($L'_{CO}(5-4)/L'_{CO}(1-0)$) ratios within the same sample and we convert the CO(1-0) LF into the corresponding CO(3--2) and CO(5--4) the variation of the bright-end can be $>1\,\rm dex$.

\begin{figure}
\centering
\includegraphics[scale=0.4]{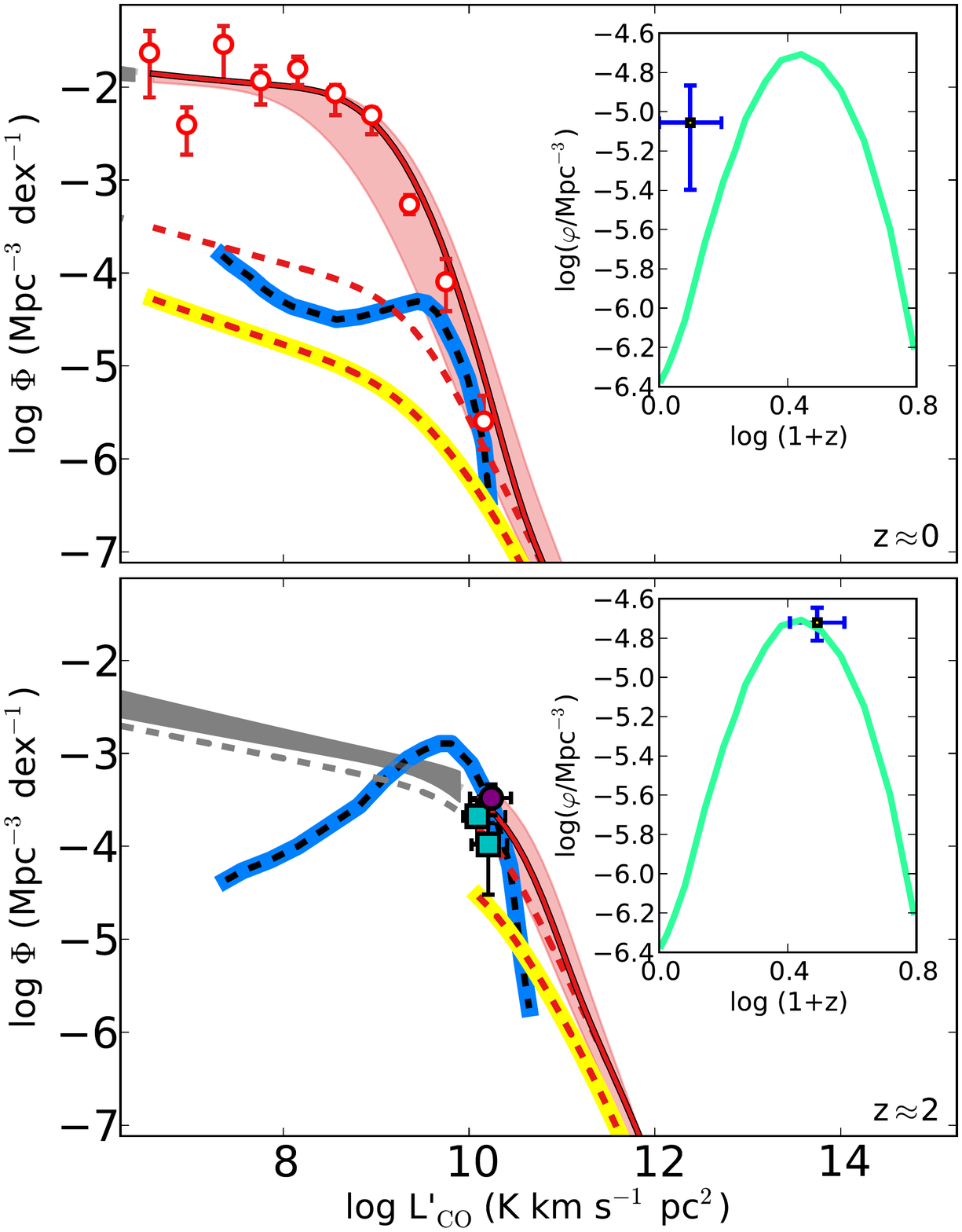}
\caption{CO(1--0) LF at $z\approx0,2$ as
  predicted by our model. The blue dashed lines indicate the
  contribution AGN-dominated galaxies ($L_{X}(2-10\,\rm{keV})>10^{44}\,\rm{erg\,s^{-1}}$) by
  \citet{lagos2012}, while the red dashed lines represent our results for the sum of \texttt{AGN1, AGN2} and \texttt{AGN-SB}. The resulting LF, up to the luminosity limit, corrected for the fraction of AGN with $L_{X}>10^{44}\,\rm{erg\,s^{-1}}$, is highlighted in yellow. In the upper/lower inset we plot in blue the number density ($\varphi$) resulting from the integration of the corrected \texttt{AGN1, AGN2} and \texttt{AGN-SB} LFs up to their luminosity limit at $z\approx0$ and $z\approx2$ respectively. The cyan solid line represents $\varphi$ for AGN with $44<\rm {log}(L_x/\rm{erg\,s^{-1}})<45$ by \citet{aird2015}.}\label{co_agn}
\end{figure}

We now exploit the potential of the LF decomposition by \citet{gruppioni2013} to separately analyze the relative contribution to the total CO LF of galaxies populations with a significant AGN activity (i.e. \texttt{AGN1, AGN2, AGN-SB}).
The contribution of these classes is shown with red dashed lines in Fig. \ref{co_agn}. 
In the same figure, blue dashed lines indicate the theoretical prediction by
\citet{lagos2012} for galaxies that host bright AGN
($L_{X}(2-10\,\rm{keV})>10^{44}\,\rm{erg\,s^{-1}}$). \citet{lagos2012} found that bright AGN are responsible for most of
the evolution with $z$ of the bright-end of the CO LF.

However, the classification proposed by \citet{lagos2012} is not directly comparable to the one adopted in
\citet{gruppioni2013} to identify AGN-dominated galaxies. As a matter of fact, in
\citet{gruppioni2013} the selection is based on the typical SED of the
objects composing the class regardless of the intrinsic $L_{X}$ of these galaxies. 
Thus, to make a more meaningful comparison, we estimate the fraction of \texttt{AGN1, AGN2, AGN-SB} with $L_{X}>10^{44}\,\rm{erg\,s^{-1}}$. To this purpose, for the sources detected in X-rays, we use the measured $L_{X}$, while for the undetected ones, we convert the total intrinsic luminosity of the accretion disk of AGN (bolometric luminosity; $L_{bol}$) into the corresponding
X-ray (2-10 keV) luminosity through the
X-ray bolometric correction $L_{X}=L_{bol}/k_{bol}(L_{bol})$ as in \citet{marconi2004}. The $L_{bol}$ has been computed by \citet{delvecchio2014} through a SED decomposition analysis performed on the same sample considered in \citet{gruppioni2013}.
\begin{figure}
\centering
\includegraphics[scale=0.35]{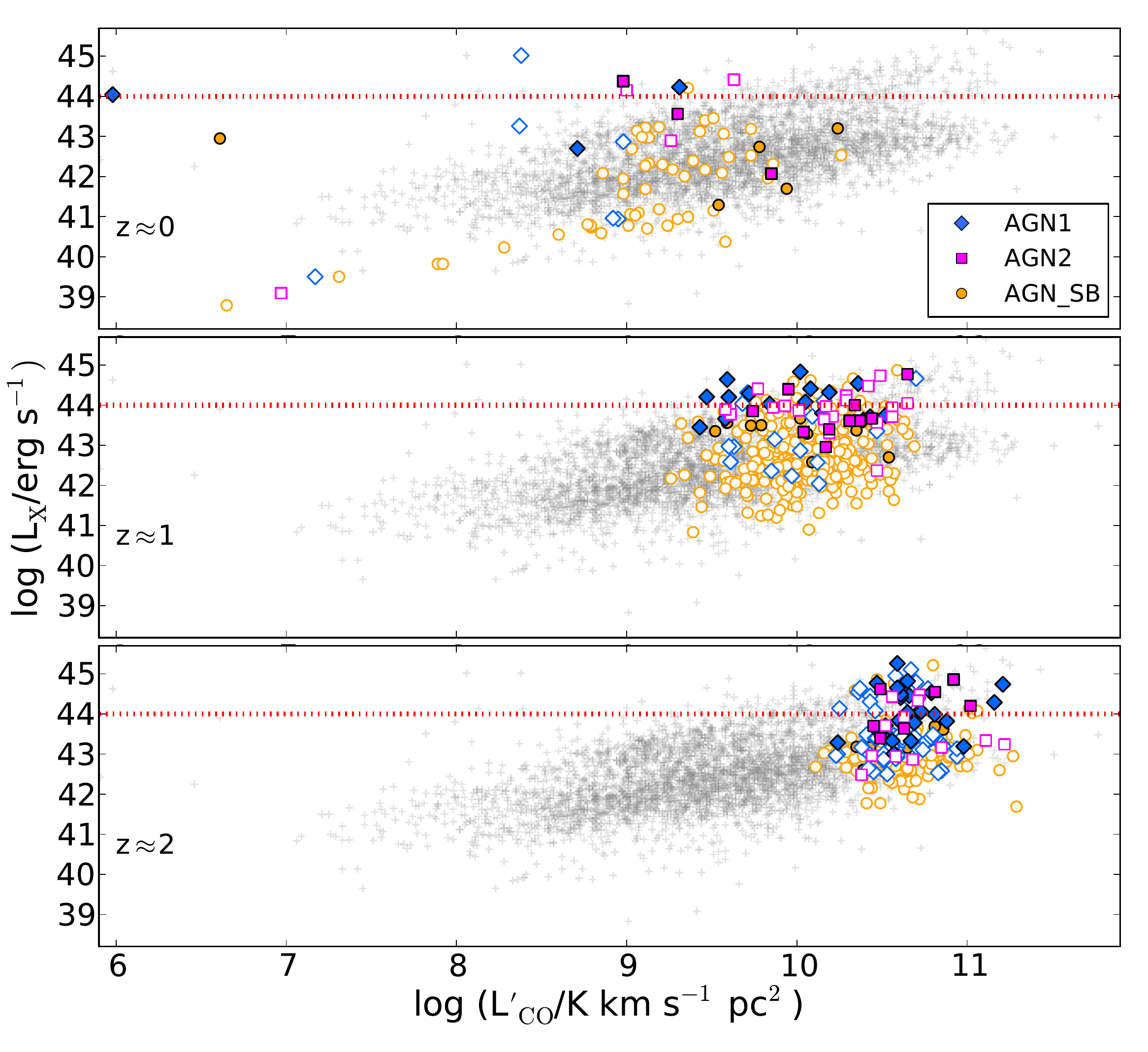}
\caption{$L_{X}$ as a function of the CO(1--0) luminosity for \texttt{AGN1} (blue diamonds) \texttt{AGN2} (magenta squares) and \texttt{AGN\_SB} (orange circles) at $z\approx0,1,2$ (upper/middle/lower panel). Filled symbols represent objects for which the X-ray luminosities are \textit{measured}, while open symbols are used for sources that are non-detected in X-ray whose $L_{X}$ is computed through the $L_{bol}$. For reference, the whole sample at all redshifts, including the \texttt{spirals, starburst, AGN\_GAL} is plotted with gray crosses. We highlight with a red dotted line the limit $L_{X}=10^{44}\,\rm{erg\,s^{-1}}$.}
\label{x-co-corr}
\end{figure}
In Fig. \ref{x-co-corr} we plot the X-ray luminosity, as obtained by the procedure described above, as a function of the $L'_{\rm CO}$ for \texttt{AGN1, AGN2}
and \texttt{AGN-SB}. We note that the points align along two parallel sequences: the higher one is populated mainly by sources that, according to the SED decomposition by \citet{gruppioni2013}, are AGN dominated (\texttt{AGN1, AGN2}), while the lower one is characterized by galaxy dominated objects (\texttt{AGN\_SB}).
Moreover, even though $L_{bol}$ and $L'_{CO}$ are not completely independent, being both indirectly based on the SED fit, no strong correlation is found between these two quantities within the same redshift range.
This implies that the correction factor that we must apply to our LF in order to match with the \citet{lagos2012} criterium can be assumed to be constant over the interval over which we have data. However, it is impossible to apply any correction to the faint end extrapolation of
the CO LF because in that range we have no clues on the actual value of the $L_{bol}$ (and thus of the $L_{X}$) with respect to the IR luminosity. 

The correction factor ($f_{X}$) is calculated by considering the Cumulative Distribution Function (CDF) of the $L_{X}$ within the \texttt{AGN1, AGN2} and \texttt{AGN-SB} classes. 
We find that at $z\approx0$ the fractions of \texttt{AGN1, AGN2} and
\texttt{AGN-SB} with $L_{X}>10^{44}\,\rm{erg\,s^{-1}}$ are respectively $f_{X}=[0.3,
0.4,0.02]$ while at $z\approx2$ they are $f_{X}=[0.4, 0.3, 0.09]$. The resulting CO LFs after correcting by the $f_{X}$ the bins in
which we have data, are highlighted in yellow in Fig. \ref{co_agn}. We note that, once we correct our sample for the $L_{X}$ threshold, our predictions are substantially lower than the theoretical one by \citet{lagos2012}. 
In principle, a possible way to explain this discrepancy would be
that some of the sources identified by
\citet{gruppioni2013} as dominated by star formation (\texttt{spiral, starburst, AGN-GAL}) could instead host AGN as powerful as
$L_{X}>10^{44}\,\rm{erg\,s^{-1}}$. However we verified that $f_X = 0$ for the objects in these classes. To further test the consistency of our predictions we calculate the number density of AGN with $L_{\rm X}>10^{44}$ erg~s$^{-1}$ by integrating up to the luminosity limit the corrected \texttt{AGN1, AGN2, AGN\_SB} LFs, and we compare the result with the number density ($\varphi$) of objects with $44<{\rm log}(L_x/{\rm erg\,s^{-1}})<45$ found by \citet{aird2015} using X-ray surveys, achieving a good consistence at $z\approx2$ but a higer value, although with a large uncertainty, at $z\approx0$ (see Fig. \ref{co_agn}). This implies that our predicted LF, although significantly lower than that of \citet{lagos2012}, yields a number density larger with respect to the number density of AGN with $L_{\rm X}>10^{44}$ erg~s$^{-1}$ derived from X-ray data. Note that the higher value of $\varphi$ found at $z\approx0$ from our infrared based work might be due to an underestimate either of the Compton thick fraction or of the obscuration correction for some X-ray sources (i.e., classified as $L_{\rm X}<10^{44}$ erg~s$^{-1}$ although intrinsically brighter).
After having excluded the hypothesis of missing a substantial fraction of
objects with $L_{\rm X}>10^{44}$ erg~s$^{-1}$ because accounted in other
classes, we derived the 3-$\sigma$ upper limits on the rest-frame
2--10~keV luminosities for the objects in the \citet{aravena2012} sample
using the XMM-{\it Newton} data.
To this purpose, we adopted the 0.5--2~keV sensitivity map, which takes into
account the effects of vignetting; the choice of this band is motivated by
the high throughput of XMM-{\it Newton} at soft X-ray energies.
Assuming a powerlaw model with photon index $\Gamma=1.7$, we obtain that
all the sources have $L_{\rm X}<3.5\times10^{43}$~erg~s$^{-1}$. 
This results is in agreement with our prediction that the data points by \citet{aravena2012} should not contain $L_{X}>10^{44}\,\rm{erg\,s^{-1}}$ AGN and therefore further support our hypothesis that the bright end of the CO LF, while mainly produced by sources that are likely AGN hosts, do not show up as extremely luminous.

\section{Discussion and Conclusions}

We calculated the CO(1--0), CO(3--2), and CO(5--4) LFs at $z\approx0,1$ and $2$ by starting from the redshift evolution of the state-of-the-art IR luminosity function presented in \citet{gruppioni2013}. We obtain the CO LF by coupling the IR luminosity function with $L_{\rm IR}-L'_{\rm CO}$ conversions specifically tailored for each class of galaxies that compose the infrared LF. Our empirical approach reproduces well the observed data/upper limits at $z\approx0$ and $z\approx2$. This is an encouraging validation of our assumption that the redshift evolution of the CO and IR LF are closely related. As a caveat, we must note that, especially at $z\approx2$, there are tensions between the predictions of the faint (and bright) end as resulting from our approach, and those obtained by SAMs. These discrepancies might be explained by several concurring reasons either on the theoretical side (e.g. difficulties in modeling feedbacks) and/or on the observational side (e.g. uncertainties in the photometric redshifts). Moreover we cannot rule out the possibility that toward $z\approx2$ and above, the evolution of IR and COs LFs might be different. Finally we demonstrate that, although AGN-dominated galaxies account for the bright end of the CO LF, at $z\approx2$ we are able to reproduce the observed points above the knee of the CO LF only if we include all the AGN dominated galaxies in our sample, regardless of their X-ray ($2-10\,\rm{keV}$) luminosity. More precisely, AGN with $L_{X}>10^{44}\,\rm{erg\,s^{-1}}$ that have been predicted by previous studies to completely account for the bright end of the CO LF are too rare to reproduce the actual CO luminosity function at $z\approx2$.

\section*{Acknowledgements}
We are grateful to G. Popping for having provided the CO luminosity functions as resulting from his work in preparation. We thank the anonymous referee for her/his insightful comments that improved significantly the paper.
\label{lastpage}
\bibliographystyle{mnras}
\bibliography{bibliography}
\end{document}